\documentclass[11pt]{article}
\usepackage[top=1in, bottom=0.7in,left=1in, right=1in]{geometry}
\usepackage{amssymb,amsmath}
\usepackage{times}
\usepackage{amsthm}
\usepackage[dvips]{graphicx}
\usepackage{subfigure}
\usepackage{float}
\usepackage{setspace}
\usepackage[centerlast,small,bf]{caption}
\newtheorem{thm}{Theorem}[section]
\newtheorem{lem}[thm]{Lemma}

\begin{document}
\date{}

\title{On finding 2-cuts and 3-edge-connected components in parallel}

\author{
Yung H. Tsin\footnote{School of Computer Science,  University of Windsor, Windsor, Ontario, Canada, N9B 3P4; peter@uwindsor.ca.
}\\
}

\maketitle \thispagestyle{empty}

\begin{abstract}
Given a connected undirected multigraph $G$ (a graph that may contain parallel edges),
the algorithm of~\cite{T07} finds the 3-edge-connected components  of $G$ in linear time using an innovative graph contraction technique based on a depth-first search.
In~\cite{T23}, it was shown that the algorithm can be extended to produce a Mader construction sequence for each 3-edge-connected component,  a cactus representation of the 2-cuts (cut-pairs) of each 2-edge-connected component of $G$, and the 1-cuts (bridges) at the same time.
In this paper, we further  extend the algorithm of~\cite{T07} to  generate the 2-cuts and
the 3-edge-connected components of $G$ simultaneously in linear time by performing only one depth-first search over the input graph.
 Previously known algorithms solve the two problems separately in multiple phases.

\vspace{12pt}

\end{abstract}

\doublespacing

\noindent

\section{Introduction}

  3-edge-connectivity is a graph-theoretic problem that is useful in
 a variety of apparently unrelated areas such as
  editing cluster and aligning genome in bioinformatics~\cite{De,Pat10},
  solving the $G$-irreducibility of Feynman diagram in physics and quantum
 chemistry~\cite{Co,KLEW22},
  placing monitors on the edges of a network in flow networks~\cite{CCY09},
  spare capacity allocation~\cite{LT13},
  layout decomposition for multiple patterning lithography~\cite{FCC12,KY13},
 and
  the traveling salesman problem~\cite{IR19}.
 The problem calls for determining the 2-cuts and the 3-edge-connected components (abbreviated 3ecc)
 of a  connected undirected graph.
  A number of linear-time algorithms for 3-edge-connectivity
  have been proposed~\cite{GI91,NI92,TWO92,T07,T09}.
 An empirical experiential study reported in~\cite{Nor07} shows that~\cite{T09} has the
 best performance in determining the 2-cuts while~\cite{T07}  has the
 best performance in determining the 3eccs.
 Both algorithms are simple in that they traverse the input graph once to accomplish their task.
 Notice that~\cite{T07} does not determine the 2-cuts
 while ~\cite{T09} needs to make another traversal over the graph after the 2-cuts are removed and
 some new edges are added to determine the 3eccs.
Recently, Georgiadis et al.~\cite{GGIK21} presented an algorithm for finding the 2-cuts
and conducted an experimental study showing that its empirical performance outperforms that of~\cite{T09}.
 They did not address the 3ecc problem.
In this article, we show that the 3ecc algorithm of~\cite{T07} can be easily modified to determine
the 2-cuts in parallel with the 3eccs by traversing the input graph only once.


\section{Basic definitions}
The definitions of the graph-theoretic concepts used in this article are standard and
  can be found in references such as~\cite{Ev, Tar72}.
  We only give some important definitions below.

An undirected graph is represented by $G=(V,E)$, where $V$ is the vertex set and $E$ is the edge set.
An edge $e$ with $u$ and $v$ as end-vertices is represented by $e=(u,v)$.
The graph may contain parallel edges (two or more edges sharing the same pair of end vertices).
The \emph{degree} of a vertex $u$ in $G$, denoted by $deg_G(u)$, is the number of edges with $u$ as an end-vertex.
A \emph{path} $P$ in $G$ is a sequence of alternating vertices
   and edges,
   $u_0 e_1 u_1 e_2 u_2 \ldots e_k u_k$, such that $u_i \in V, 0 \le i
   \le k,$ $e_i  = (u_{i-1}, u_i), 1 \le i \le k$, where $u_i, 0 \le i
   \le k,$ are distinct with the exception that $u_0$ and $u_k$ may be identical.
    The edges $e_i, 1 \le i \le k,$ could be omitted if no
    confusion could occur.
     The path is a \emph{null path} if $k=0$ and is a \emph{cycle} if  $u_0 = u_k$.
     The path is called an $u_0-u_k$ \emph{path} with vertices $u_0$ and $u_k$ as \emph{terminating vertices}
and $u_i, 1 \le i <k$, as \emph{internal} vertices.
    If the path $P$ is given an orientation from $u_0$ to $u_k$, then $u_0$ is the \emph{source}, denoted by $s(P)$, and $u_k$ is the \emph{sink}, denoted by $t(P)$, of $P$ and the path $P$ is also represented by $u_0 \rightsquigarrow_G u_k$.
         The graph $G$ is \emph{connected} if $\forall u,v \in V$, there is a $u-v$ path
     in it. It is \emph{disconnected} otherwise.
     Let $G = (V,E)$ be a connected graph.
  An edge is a 1-$cut$ (or \emph{bridge}) in $G$ if removing it from $G$ results in a
  disconnected graph.
   The graph $G$ is \emph{2-edge-connected}  if it has no 1-cuts.
   A 2-$cut$ (or \emph{cut-pair}) of $G$ is a pair of edges whose removal
   results in a disconnected graph and neither edge is a bridge.
   A \emph{cut-edge} is an edge in a cut-pair.
     $G$ is \emph{3-edge-connected} if it does not have 1-cut or 2-cut.
     A \emph{3-edge-connected component} (abbreviated $3ecc$) of $G$ is a maximal subset $U \subseteq V$ such that $\forall u, v \in U, u \neq v$, there exists three edge-disjoint $u - v$ paths in $G$.
 A graph $G' =(V', E')$ is a \emph{subgraph} of $G$ if $V' \subseteq V$ and $E' \subseteq E$.
  Let $U \subseteq V$, the \emph{subgraph of $G$ induced by $U$}, denoted by $G_{\langle U \rangle}$, is the maximal subgraph of $G$ whose vertex set is $U$.
     Let $D \subseteq E$, $G \setminus D$ denotes the graph resulting from $G$ after
     the edges in $D$ are removed.

   It is well-known that performing a depth-first search~\cite{Tar72} (henceforth, abbreviated $\mathit{dfs}$) over $G$ creates a spanning tree $T=(V,E_T)$, called $\mathit{dfs}$ \emph{spanning tree} of $G$.
    $T$ is a \emph{rooted tree} with $r$, the vertex at which the $\mathit{dfs}$ begins, as the \emph{root}.
   Every vertex $u$ is assigned a distinct integer, $\mathit{dfs}(u)$, called its \emph{dfs number}, which is its rank of $u$ in the order the vertices are visited by the $\mathit{dfs}$ for the first time.
   The edges in $T$ are called \emph{tree edge}s and the edges not in $T$ are called \emph{back-edge}s.
   For $y \in V \setminus \{r\}$, there is a unique tree-edge $(x,y)$ such that $\mathit{dfs}(x) < \mathit{dfs}(y)$.
   Vertex $x$ is called the \emph{parent} of  $y$, denoted by $parent(y)$, while $y$ is a \emph{child} of  $x$.
     The tree-edge is called the \emph{parent edge} of $y$ and a \emph{child edge} of $x$
     and is denoted by $x \rightarrow y$ or $y \leftarrow x$.
     If  $e = (x,y)$, we may write $x \overset{e}{\rightarrow} y$ or $y \overset{e}{\leftarrow} x$.
     A back-edge $(x,y)$ with $\mathit{dfs}(y) < \mathit{dfs}(x)$,
     is an \emph{outgoing back-edge} (\emph{incoming back-edge}, resp.) of $x$
     ($y$, resp.). The back-edge is denoted by
  $y \curvearrowleft x$ or $x \curvearrowright y$,
     where $x$ is the \emph{tail} and $y$ is the \emph{head}.
  If $e = (x,y)$, we may write
  $y \overset{e}{\curvearrowleft} x$ or  $x \overset{e}{\curvearrowright} y$.

\noindent $\forall w \in V, lowpt(w)  = \min (  \{dfs(w)\} \cup \{ dfs(u) \ | \ (u \curvearrowleft w) \in E \setminus E_T \}  \cup \{ lowpt(u) \ | \ (w
\rightarrow u) \in  E_T \}  )$.
Specifically, $lowpt(w)$ is the smallest $\mathit{dfs}$ number of a  vertex reachable from $w$ via a (possibly null)  tree-path followed by a back-edge~\cite{Tar72}.

    A path connecting a vertex $x$ with a vertex $y$ in the rooted $T$
    is denoted by $x \rightsquigarrow_T y$.
      A vertex $v$ is an \emph{ancestor} of a vertex $u$, denoted by $v \preceq u$, if and only if it $v$ lies on $r \rightsquigarrow_T u$.
      Vertex $v$ is a \emph{proper ancestor} of vertex $u$, denoted by $v \prec u$, if
      $v \preceq u$ and $v \neq u$.
      Vertex $u$ is a (proper) \emph{descendant} of vertex $v$  if and only if
      $v$ is a (proper) ancestor of vertex $u$.
      The \emph{subtree} of $T$ rooted at $u$, denoted by $T_u$, is the subgraph of $T$ induced by the descendents of $u$.
The notation
  $x \rightarrow y \rightsquigarrow_T z$ represents a path consisting of $(x \rightarrow y)$ and $(y \rightsquigarrow_T z)$. The notations
  $x \curvearrowright y \rightsquigarrow_T z$,
  $x \rightsquigarrow_T y \curvearrowright z$, etc. are defined similarly.
If edge $(x,y)$ is a tree-edge with $x$ as the parent, we use $(x,y), (x \rightarrow y)$ and $(y \leftarrow x)$ interchangeably.
If edge $(x,y)$ is a back-edge with $x$ as the tail, we use $(x,y), (x \curvearrowright y)$ and $(y \curvearrowleft x)$ interchangeably.

\section{Finding 2-cuts}

The following are well-known facts about 2-cuts.

\begin{lem}\label{BasicFacts}
  \emph{\textbf{\cite{TWO92,T09}}}
  Let $G=(V,E)$ be a 2-edge-connected graph and $T=(V,E_T)$ be a $\mathit{dfs}$ tree of $G$ with $r$ as the root. Let $\{e,e'\}$ be a 2-cut of $G$.

\vspace{-9pt}
\begin{description}
  \item[$(i)$] At most one of $e$ and $e'$ is a back-edge;

\vspace{-9pt}
  \item[$(ii)$] if both $e$ and $e'$ are tree-edges, then $e$ and $e'$ lie on a tree-path connecting the root $r$ to a leaf;

\vspace{-9pt}
  \item[$(iii)$] if $y \overset{e}{\curvearrowleft} x \in E \setminus E_T$ and $w \overset{e'}{\rightarrow} u \in E_T$, then $y \preceq w$ while $u \preceq  x$ in $T$;

\vspace{-9pt}
  \item[$(iv)$] if $\{e',e''\}$ is a 2-cut, then $\{e,e''\}$ is also a 2-cut.

\end{description}
\end{lem}

  As an immediate consequence of Lemma~\ref{BasicFacts}$(iv)$,
  the set of all cut-edges can be partitioned into a collection of disjoint subsets
$\mathcal{E}_i, 1 \le i \le \eta,$ called \emph{cut-edge chains},
 such that every two cut-edges in the same subset form a 2-cut and no two edges from different subsets form a 2-cut.
  By  Lemma~\ref{BasicFacts}$(i)$, each $\mathcal{E}_i$ contains at most one back-edge.
  A cut-edge chain containing a back-edge is a  \emph{TB-cut-edge chain}.
  A cut-edge chain containing only tree-edges is a \emph{TT-cut-edge chain}.

 Let $\mathcal{E}_i$ be a cut-edge chain.
 By Lemma~\ref{BasicFacts}$(ii)$, the edges in $\mathcal{E}_i$ can be lined up along a root-to-leaf path in $T$ as follows:
\begin{description}
  \item[$(i)$]
  If $\mathcal{E}_i$ is of \emph{TT} type,
  the edges in $\mathcal{E}_i$ can be lined up along a root-to-leaf path in $T$
    in an order $e_1 e_2 \ldots e_{|\mathcal{E}_i|}$, where
    $x_j \overset{e_j}{\rightarrow} y_j, 1 \le j \le |\mathcal{E}_i|$,
 such that $y_{j+1} \preceq x_j, 1 \le j < |\mathcal{E}_i|$ (Figure~1$(i)$).

  \item[$(ii)$]
  If $\mathcal{E}_i$ is of \emph{TB} type,
  let the back-edge be $y_1 \overset{e_1}{\curvearrowleft} x_1$. Then,
    the tree-edges in $\mathcal{E}_i$ can be lined up along a root-to-leaf path in $T$
    in an order $e_2 e_3 \ldots e_{|\mathcal{E}_i|}$, where
    $x_j \overset{e_j}{\rightarrow} y_j, 2 \le j \le |\mathcal{E}_i|$,
 such that $y_{j+1} \preceq x_j, 2 \le j < |\mathcal{E}_i|$ and
    $y_1 \preceq x_{|\mathcal{E}_i|}, y_2 \preceq x_1$ (Figure~1$(ii)$).

\end{description}

\begin{figure}\label{fig1}
\centering
\includegraphics[width=5in]{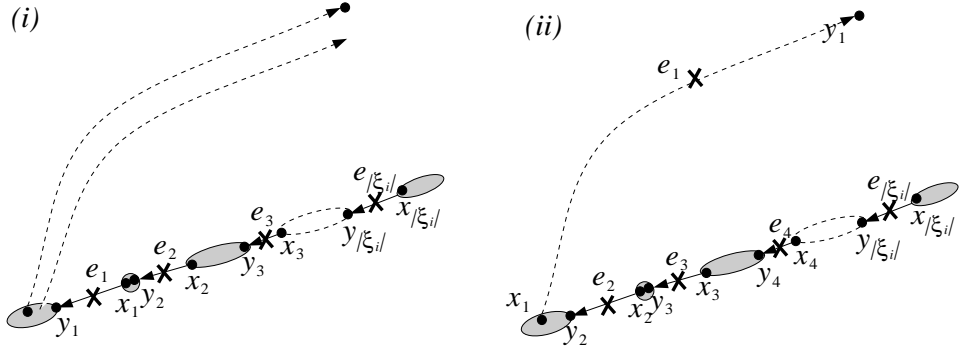}
\caption[ ]{$(i)$ a \emph{TT}-cut-pair chain;   {$(ii)$ a \emph{TB}-cut-pair chain}.}
\end{figure}

  The cut-edge $e_1$ is called the \emph{generator} of $\mathcal{E}_i$~\cite{TWO92,T09}.
  Hence, finding all the 2-cuts in $G$ can be reduced to determining all the cut-edge chains of $G$.

   Before explaining how to modify
   \textbf{Algorithm} \texttt{3-edge-connectivity} of~\cite{T07} to generate the  cut-pair chains
   $\mathcal{E}_i, 1 \le i \le \eta$, we give the algorithm a brief review.

The key idea underlying the algorithm is to use a graph contraction operation, called \emph{absorb-eject}, to gradually transform the input graph $G$ into a \emph{null} (i.e., edgeless) graph
of which each vertex corresponds to a distinct 3-edge-connected component of $G$
during a depth-first search.

\noindent \textbf{Definition:}
  Let $G' = (V', E')$ and $e=(w,u) \in E'$ such that:

$(i)$ $deg_{G'}(u) = 2$, or
$(ii)$ $e$ is not a cut-edge (which implies $deg_{G'}(u) \neq 2$).

   Applying the \emph{absorb-eject} operation to $e$ at $w$ results in the graph $G'/e = (V'', E'')$, where

\hspace{20pt}   $V'' = \left\{
           \begin{array}{ll}
             V',                 & \hbox{if $deg_{G'}(u) = 2$;} \\
             V' \setminus \{u\}, & \hbox{if $e$ is not a cut-edge, }
           \end{array}
         \right.
  $

\vspace{6pt}
\hspace{20pt}  $E'' = E' \setminus E_u \cup E_{w^+}$,
where $E_u$ is the set of edges incident on $u$ in $G'$ and
      $E_{w^+} = \{ f' = (w,z) \mid \exists f = (u, z) \in E_u$, for some $z \in V' \backslash \{w\} \}$.~(Figure~2)

The edge $f'=(w,z)$ is called an embodiment of the edge $f = (u,z)$.
In general, an \emph{embodiment} of an edge $f$ is the edge $f$ itself, or
an edge created to replace $f$ as a result of applying the absorb-eject operation,
 or an embodiment of an embodiment of $f$.

 \begin{figure}
 \centering
 \includegraphics[width=4.5in]{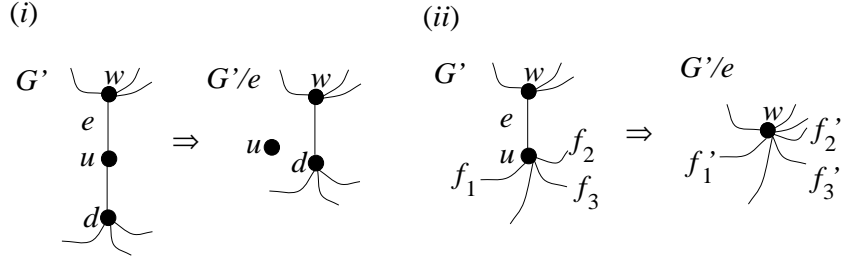}
 \caption[ ]{The absorb-eject operation: $(i)$ $deg_{G'}(u) = 2$;
 $(ii)$ edge $(w,u)$ is not a cut-edge. }
 \end{figure}\label{Fig3-0.eps}

Starting with the input graph $G = (V, E)$, using the absorb-eject operation, the graph is
gradually transformed so that vertices that have been confirmed to be belonging to the same $3ecc$ are merged into one vertex, called a \emph{supervertex}.
Each supervertex is represented by a vertex $w \in V$ and a set $\sigma(w) (\subseteq V)$
consisting of vertices that have been confirmed
to be belonging to the same $3ecc$
as $w$. Initially, each vertex $w$ is regarded as a supervertex with $\sigma(w) = \{w\}$.
When two adjacent supervertices $w$ and $u$ are known to be belonging to the same
$3ecc$, the absorb-eject operation is applied to have one of them, say $w$, absorbing the other resulting in $\sigma(w) := \sigma(w) \cup \sigma(u)$.
When a supervertex containing all  vertices of a $3ecc$ is formed,
it will become \emph{of degree one or two} in the transformed graph (corresponding to a 1-cut or a 2-cut is found)\footnote{In~\cite{T07}, it is pointed out that \textbf{Algorithm} 3-edge-connectivity can be easily modified to handle non-2-edge-connected graphs.}.
 Then, the absorb-eject operation is applied to an adjacent supervertex
 to separate (\emph{eject}) the supervertex from the graph making it an isolated vertex.
At the end, the graph is transformed into a collection
of isolated supervertices each of which consists of the vertices of a distinct $3ecc$ of $G$.

During the depth-first search, at each vertex $w \in V$,
when the adjacency list of $w$ is completely processed,
let $\hat{G}_w$ be the graph to which $G$ has been transformed at that point in time.
The subtree of $T$ rooted at $w$, $T_w$, has been transformed into a path of supervertices in $\hat{G}_w$,
 $\mathcal{P}_w: (w=)w_0 w_1 w_2 \ldots w_k$, called the $w$-\emph{path}, and a set of isolated supervertices each of which corresponds to a $3ecc$ residing in $T_w$.
   The $w$-path has the properties summarised in the following lemma.

\vspace{-3pt}
\begin{lem} \label{Tsin07}

\emph{[Lemma 6 of~\cite{T07}]} ~
    Let $\mathcal{P}_w: (w=)w_0 w_1 w_2 \ldots w_k$ and
    $\hat{G}_w$ is the graph to which $G$ has been transformed at that point in time                      \emph{(}Figure~3\emph{)}.

 \begin{description}
    \vspace{-6pt}
   \item[$(i)$] $deg_{\hat{G}_w}(w_0) \ge 1$  and $deg_{\hat{G}_w}(w_i) \ge 3, 1 \le i \le k$;

   \vspace{-6pt}
   \item[$(ii)$]  for each back-edge $f = (w_i \curvearrowright x), 0 \le i \le k,$ in $\hat{G}_w$,
      $x \preceq v $ (i.e. $x$ lies on the $r \rightsquigarrow_T v$ tree-path);

    \vspace{-6pt}
   \item[$(iii)$] $\exists (w_k \curvearrowright z)$ in $\hat{G}_w$ such that $dfs(z) = lowpt(w)$. 
 \end{description}
\end{lem}

\vspace{-6pt}
 Specifically, on the $w$-path, the degree of every supervertex is at least \emph{three} except that of $w_0$,
 there is no back-edge connecting two supervertices on the $w$-path,
and the  supervertex $w_k$ has an outgoing back-edge reaching the vertex whose \emph{dfs} number is $lowpt(w)$.

 The $w$-path is constructed as follows.
 Initially, the $w$-path is the null path $w$ which is the \emph{current} $w$-\emph{path},
 and $lowpt(w) = dfs(w)$.
 When the \emph{dfs} backtracks from a child $u$, let
 $\mathcal{P}_u: (u=)u_0 u_1 u_2 \ldots u_h$ be the $u$-\emph{path},
 and
 $\hat{G}_u$ be the graph to which $G$ has been transformed at that point in time.

 If $deg_{\hat{G}_u}(u) = 1$, then $(w,u)$ is a 1-cut and $\sigma(u)$ is a $3ecc$ of $G$.
 Edge  $(w,u)$ is removed making $u$  an isolated supervertex.
 The current $w$-path remains unchanged.

 If $deg_{\hat{G}_u}(u) = 2$, then $\{(w \rightarrow u), (u \rightarrow u_1)\}$ or
  $\{(w \rightarrow u), (z \curvearrowleft u)\}$,
  where $dfs(z) = lowpt(u)$, is a 2-cut
 implying $\sigma(u)$ is a $3ecc$ of $G$.
   The absorb-eject operation is applied to edge $(w,u)$ at $w$ to eject supervertex $u$ from the $u$-path making it an isolated supervertex and the
   $u$-path is shorten to $u_1 u_2 \ldots u_k$ in the former case  or vanishes in the latter case.
 Furthermore,
   $(a)$ if $lowpt(w) \le lowpt(u)$,
   then the vertices in the supervertices on the $u$-path must all belong to the same $3ecc$ as $w$.
 The supervertices are thus absorbed by $w$ through a sequence of absorb-eject operations.
 $(b)$ if $lowpt(w) > lowpt(u)$, then the vertices in the supervertices on the current $w$-path must all belong to the same $3ecc$ as $w$;
 the supervertices are thus absorbed by $w$.
 Moreover, $lowpt(w)$ is updated to $lowpt(u)$,
and  the $u$-path after extended to include $(w,u)$ becomes the current $w$-path, if $u$ is not ejected.
 If $u$ is ejected, then the $u$-path after extended to include $(w,u_1)$ becomes the current $w$-path, if   $u_1$ exists, and is the null path $w$, otherwise.

   When an outgoing back-edge of $w$, $(z \curvearrowleft w)$, with $dfs(z) < lowpt(w)$
is encountered,
vertex $w$ absorbs the current $w$-path because all the supervertices on it belong to the same $3ecc$ as $w$; $lowpt(w)$ and the current $w$-path are then updated to $dfs(z)$ and the null path $w$, respectively.

When an incoming back-edge of $w$, $(w \curvearrowleft x)$, is encountered,
if $\exists h, 1 \le h \le k,$ such that $x \in \sigma(w_{h})$ on the current $w$-path,
the vertices in $\sigma(w_i), 1 \le i \le h$,
  must all belong to the same $3ecc$ as $w$.
   The supervertices $w_i, 1 \le i \le {h},$ are thus absorbed by $w$
   (Property $(ii)$ of the $w$-path thus holds) and
the current $w$-path is shortened to $w w_{h+1} w_{h+2} \ldots w_k$.

When the adjacency list of $w$ is completely processed,
if $w \neq r$, the current $w$-path becomes the $w$-path,
and
the depth-first search backtracks to the parent vertex of $w$.
Otherwise, the input graph $G$ has been transformed into a collection of isolated supervertices each of which contains the vertices of a
distinct $3ecc$ of $G$.
A complete example is given in~\cite{T07}, pp.132-133.


During an execution of the algorithm,
when the depth-first search backtracks from a vertex $u$ to the parent vertex $w$,
if $deg_{\hat{G}_u}(u) = 2$, then $\sigma(u)$ is a 3ecc, and $\{(w \rightarrow u),(u \rightarrow u_1)\}$ or $\{(w \rightarrow u), (u \curvearrowright z)\}$ is the corresponding 2-cut.
Unfortunately, the 2-cut is not in $G$, but in the transformed graph $\hat{G}_u$.
Specifically,
  the corresponding 2-cut in $G$ is $\{(w,u), ( x, y)\}$, such that $(u \rightarrow u_1)$ or $(u \curvearrowright z)$ is an embodiment of $( x, y)$.
We shall show that the cut-edge $( x, y)$ can be easily determined as follows:

  $(i)$ if the 2-cut in $\hat{G}_u$ is $\{(w \rightarrow u),(u \rightarrow u_1)\}$,
  the corresponding 2-cut in $G$ is $\{(w,u), (parent(u_1), u_1)\}$.

  $(ii)$ if the 2-cut in $\hat{G}_u$ is $\{(w \rightarrow u),  (u \curvearrowright z )\}$,
   the corresponding 2-cut in $G$ is $\{(w,u), (x,z)\}$, where $x$ is the unique vertex
   such that $u \preceq x$ and
    $lowpt(x) = lowpt(u) = dfs(z)$.

\begin{lem}\label{2cut-iff-1}
 \emph{\textbf{\cite{TWO92,T09}}}
Let $e, e' \in E$ be such that $e=(w \rightarrow u),  e'=(x,y)$, and $dfs(u) \le dfs(x)$.

\vspace{-9pt}
\begin{description}
  \item[$(i)$] if $e'$ is a tree-edge $(x \rightarrow y)$, then $\{e, e'\}$ is a 2-cut in $G$ if and only if
   there does not exist a back-edge $(s \curvearrowright t)$ such that either;

(a) $s$ is a descendant of $u$ and not of $y$ while $t$ is an ancestor of $w$, or

(b) $s$ is a descendant of $y$ while $t$ is a descendant of $u$ and not of $y$.

\vspace{-6pt}
  \item[$(ii)$] If $e'$ is a back-edge $(y \curvearrowleft x)$, then $\{e, e'\}$ is a 2-cut in $G$ if and only if
  there does not exist a back-edge $f = (s \curvearrowright t)$ such that $f \neq e'$ and $s$ is
a descendant of $u$ while $t$ is an ancestor of $w$.

\end{description}

\end{lem}

\begin{lem}\label{2cut-iff-2}

During the depth-first search,  when the search backtracks from a vertex
$u$ to its parent vertex $w$,
Let $\mathcal{P}_u: (u=) u_0 u_1 u_2 \ldots u_k$ be the $u$-path.

If $deg_{\hat{G}_u}(u) = 2$, and

\vspace{-9pt}
\begin{description}
  \item[$(i)$] $\{(w \rightarrow u), (u \rightarrow u_1)\}$ is the 2-cut.
   Then the corresponding 2-cut in $G$ is $\{(w,u), (parent(u_1), u_1)\}$;

\vspace{-6pt}
  \item[$(ii)$]  $\{(w \rightarrow u), (u \curvearrowright z)\}$ is the 2-cut (i.e., $k = 0$).
    Then the corresponding 2-cut in $G$ is
    $\{(w,u), (x \curvearrowright z)\}$, where $(x \curvearrowright z)$ is the unique edge such that
    $u \preceq x$, and $lowpt(x) = lowpt(u) = \mathit{dfs}(z)$.

\end{description}

\end{lem}

\vspace{-9pt}
\noindent \textbf{Proof:}

\vspace{-9pt}
\begin{description}
  \item[$(i)$]
  The $u$-path contains $u$ and $u_1$ implies that there is a $u$-$u_1$ path in $G$.
  Since $u_1$ remains on the $u$-path when the depth-first search backtracks from $u$ to $w$,
  there does not exist a back-edge $(s \curvearrowright t)$ such that
  $s$ is a descendant of $u_1$ while $t$ is a descendant of $u$ and not of $u_1$;
  otherwise, let $(s \curvearrowright t)$ be one that has $t$ closest to $u_1$,
  then $u_1$ would have been absorbed by vertex $t$ when $t$ became the current vertex of the
  $dfs$.
  Similarly, there does not exist a back-edge $(s \curvearrowright t)$ such that
  $s$ is a descendant of $u$ and not of $u_1$ while $t$ is an ancestor of $w$;
  otherwise, $u$ would have an incident edge which is an embodiment of $(s \curvearrowright t)$ when
  the depth-first search backtracks from $u$ to $w$,
  implying $deg_{\hat{G}_u}(u) > 2$, which contradicts the assumption $deg_{\hat{G}_u}(u) = 2$.
  Hence, by Lemma~\ref{2cut-iff-1}$(i)$,
  $\{(w,u), (parent(u_1), u_1)\}$ is the 2-cut corresponding to $\sigma(u)$ in $G$.

\vspace{-6pt}
  \item[$(ii)$]
  Since $u = u_k$, by Lemma~\ref{Tsin07}$(iii)$,  $dfs(z) = lowpt(u)$.
  Let the 2-cut in $G$ corresponding to $\{(w \rightarrow u), (u \curvearrowright z)\}$ be $\{(w \rightarrow u), (x \curvearrowright z)\}$.
  Since  $(u \curvearrowright z)\}$ is an embodiment of $ (x \curvearrowright z)$,
  $u \preceq x$ which implies that $lowpt(u) \le lowpt(x)$.
   But $(x \curvearrowright z)$ implies $lowpt(x) \le dfs(z) = lowpt(u)$,
We thus have $lowpt(x) = lowpt(u) = dfs(z)$.
By Lemma~\ref{2cut-iff-1}$(ii)$,
there does not exist a back-edge $(s \curvearrowright t)$ such that
  $u \preceq s$ and $t \preceq w$.
   Hence, $(x \curvearrowright z)$ is the unique edge such that $u \preceq x$, and
 $lowpt(u) = lowpt(x) = \mathit{dfs}(z)$.
   \ \ \ \ \   $\blacksquare$

\end{description}

Based on Lemma~\ref{2cut-iff-2}, we determine the cut-edge that forms a 2-cut with $(w,u)$ in $G$ as follow:

\vspace{-9pt}
\begin{description}
  \item[$(i)$]
    To determine $(parent(u_1), u_1)$, we have to know the parent vertex of every vertex $u \in V \backslash \{r\}$ in $T$.
    The parent vertex of $u$ is $w$ if
 the depth-first search advances from  $w$ to $u$. This information can be stored in an array $parent[w], w \in V$, where $parent[r] = \perp$ (undefined).

\vspace{-6pt}
  \item[$(ii)$]
  During the depth-first search,
  at each vertex $x$ where $\exists (z \curvearrowleft x)$ such that $dfs(z) = lowpt(x)$,
  let $gen(x) = x$ and $low(x) = z$. Clearly, $low(x)$ can be determined in parallel with $lowpt(x)$.
  The edge $(gen(x) \curvearrowright low(gen(x))) = (x \curvearrowright z) $ is a potential \emph{TB}-cut-edge chain generator.

When a vertex $v$ absorbs a section of its current $v$-path on which $x$ is the last vertex,
$(x \curvearrowright z)$ is replaced by its embodiment $(v \curvearrowright z)$ in the resulting graph.
By letting $gen(v) := gen(x)$, $(x \curvearrowright z)$ can be retrieved as $(x \curvearrowright z) = (gen(x) \curvearrowright low(gen(x)))  = (gen(v) \curvearrowright low(gen(v)))$ if needed.
The $gen(x)$ value is transmitted upwards in $T$ this way, until a vertex $u$ is reached, where
$deg_{\hat{G}_u}(u) = 2$.
Then $\{ (w \rightarrow u),(u \curvearrowright z) \}$ is a 2-cut, where
 $(u \curvearrowright z)$ is the embodiment of $(x \curvearrowright z)$ in $\hat{G}_u$,
 Then, as $gen(u) = x$, $(x \curvearrowright z)$ can be retrieved through $(gen(u) \curvearrowright low(gen(u)))$  and
  the corresponding 2-cut in $G$ is
$\{ (w,u),$ $(gen(u) \curvearrowright low(gen(u)))\}$.
After $u$ is ejected,
if $lowpt(u) < lowpt(w)$, then $gen(w) := gen(u)$,
because $(x \curvearrowright z)$ still has the potential to form new 2-cuts with other edges in $G$.
  Otherwise,
  $gen(x)$ and $low(gen(x))$ become irrelevant as $(x \curvearrowright z)$ can no longer form 2-cuts with other edges.

\end{description}

The following is a pseudo-code of the modified algorithm. For clarity,
the new instructions for generating the 2-cuts are marked by $\bullet$.
The following variables are used to stored the cut-edge chains:

\vspace{-12pt}
\begin{itemize}
  \item $tChain(x)$: a \emph{TT}-cut-edge chain whose generator is $(parent(x) \rightarrow x)$,
\vspace{-9pt}
  \item $bChain(x)$: a \emph{TB}-cut-edge chain whose generator is $(low(x) \curvearrowleft x)$.

\end{itemize}


   When a vertex $u$ is ejected,
 \textbf{Procedure} \texttt{Gen-$\mathcal{CS}$} is called to add the cut-edge $(w,u)$  to $tChain(u_1)$ if both cut-edges are tree-edges, where $u_1$ is the vertex following $u$ in the $u$-path before $u$ was ejected, or to $bChain(gen(u))$, otherwise. In the latter case, $gen(u)$ is transferred to $gen(w)$ if $lowpt(u) < lowpt(w)$.
 \textbf{Procedure}  \texttt{Absorb-path} is called to absorb the current $w$-path or a $u$-path.
  \textbf{Procedure} \texttt{Absorb-path}
  is called when vertex $w$ absorbs  a section of the current $w$-path.

  In~\cite{T07}, $deg(w), \forall w \in V$, are calculated implicitly.
  We calculate them explicitly based on~\cite{NT14}.
For clarity,
although we include the instructions for generating the $3ecc$s,
we do not involve them in our discussion as their correctness is covered in~\cite{T07}.
we also excluded the instructions for handling parallel edges.
They can be accounted for as follows:

Initially, $\forall w \in V$,
$parent\_edge(w) := false$.
When $w$ is the current vertex of the \emph{dfs}, on encountering the edge $(w, v)$, where $v=parent(w)$, for the first time, $parent\_edge(w) := true$ and the edge is skipped.
For each subsequent $(w,v)$, $parent\_edge(w) = true$ implies that the edge is to be processed as an outgoing back-edge of $w$.

\begin{singlespacing}

%

\noindent\hrulefill

\noindent \textbf{Algorithm} \texttt{3-edge-connectivity}

\vspace{-6pt}
\noindent\hrulefill

\footnotesize

\noindent \textbf{Input:} A connected undirected multigraph $G=(V,E)$;

\noindent \textbf{Output:} $\left\{
                              \begin{array}{ll}
                                \text{The cut-edge chains of $G$;} & \hbox{     } \\
                                \text{the 3-edge-connected components of $G$;} & \hbox{     } \\
                                \text{The 1-cuts of $G$.} & \hbox{     }
                              \end{array}
                            \right.$

\noindent \textbf{begin}

   \textbf{for each} $w \in V$ \textbf{do}

\hspace{30pt}   $ deg(w) := 0; \
                  \sigma(w) := \{w\}; \
                  dfs(w) := 0; \
                  nd(w) := 1;$

\hspace{30pt}     $tChain(w) := \perp$;  \ \ \ \ \  // the \emph{TT}-cut-edge chain with $(w, parent(w))$ as   the generator

\hspace{30pt}     $bChain(w) := \perp$;  \ \ \ \ \  // the \emph{TB}-cut-edge chain with $(w, low(w))$ as the generator

\hspace{30pt}  $gen(w) := w; \  low(w) := w$;  \ \ \ \  // initialize the \emph{potential} \emph{TB}-type generator attached to $w$ with self-loop $(w \curvearrowleft w)$

\hspace{30pt}     $Bridge(w) := \emptyset$;  \ \ \ \ \  // to store the 1-cuts

  $\mathit{count} := 1$;     \ \ \ \ \  // $\mathit{dfs}$ number for next vertex

  \texttt{3-edge-connect}$(r,\perp)$;

\noindent \textbf{end.}\\

\noindent \textbf{Procedure} \texttt{3-edge-connect}$(w,v)$                    \ \ \ \ \  // $v$ is the parent of $w$

\noindent \textbf{begin}

  $dfs(w):=\mathit{count}$; \
 $\mathit{count} := \mathit{count} + 1$; \ $parent(w) := v$;

 $lowpt(w) := dfs(w)$;
\hspace{30pt}
 $next(w) := \perp;$    \ \ \ \ \ \ \ \ \ \    // initialize the $w$-path to $\mathcal{P}_w: w$

\noindent \textbf{1} \hspace{5pt} \textbf{for each} $(u$ in $L[w])$ \textbf{do}     \hspace{22pt} // scan the adjacency list of $w$

$\bullet$    \hspace{5pt}   $deg(w) := deg(w) + 1$;  \ \ \ \ \  // found a new edge incident on $w$; update $deg(w)$

\hspace{-8pt} \textbf{1.1} \hspace{5pt}
 \textbf{if}  $(dfs(u) = 0)$ \textbf{then}  \ \ \ \ \ \ \    //  $u$ is unvisited

  \hspace{42pt} \texttt{3-edge-connect}$(u, w)$;  \hspace{20pt}   // $\mathit{dfs}$ advances to $u$

$\bullet$     \hspace{35pt}
     $nd(w) := nd(w) + nd(u)$;   \hspace{20pt}   // update $nd(w)$

    \hspace{10pt}\textbf{1.1.1} \hspace{12pt}
    \textbf{if} $(deg(u) \le 2)$ \textbf{then}      
      \hspace{45pt} // found a 1-cut or 2-cut

$\bullet$    \hspace{64pt}
      \texttt{Gen}-$\mathcal{CS}(w, u, \mathcal{P}_u)$    \ \ \ \ \  \hspace{30pt} // generate the 1-cut or 2-cut

    \hspace{10pt}\textbf{1.1.2}  \hspace{12pt}
    \textbf{if} $(lowpt(w) \le lowpt(u))$ \textbf{then}

$\bullet$       \hspace{70pt}     $w' := next(w);  next(w) := u$;
                                    \hspace{10pt}  // save the current $w$-path and attach the $u$-path to $w$

               \hspace{77pt}       Absorb-path($ w $);
                                      \ \ \ \ \ \hspace{55pt}  //  $w$ absorbs the $u$-path

$\bullet$       \hspace{70pt}      $next(w) := w'$;
                                      \ \ \ \ \ \hspace{55pt}  //  restore the current $w$-path

      \hspace{45pt} \textbf{else}

           \hspace{70pt}   $lowpt(w) := lowpt(u)$;    \hspace{45pt}   //  update $lowpt(w)$

           \hspace{70pt}   Absorb-path$(w)$;
                                      \ \ \ \ \   \hspace{60pt}   // $w$ absorbs the current $w$-path

           \hspace{70pt}  $next(w) := u$; \ \ \ \ \   \hspace{65pt} // attach the $u$-path to $w$ to form the new current $w$-path

\hspace{-8pt} \textbf{1.2} \hspace{5pt}  \textbf{else if} $(dfs(u) < dfs(w))$ \textbf{then} \ \ \ \ \   // an  outgoing back-edge

    \hspace{50pt} \textbf{if} $(dfs(u) < lowpt(w))$ \textbf{then}

       \hspace{75pt}  Absorb-path$( w )$;      \hspace{85pt} // $w$ absorbs the current $w$-path

        \hspace{75pt} $lowpt(w) := dfs(u)$;    \  $next(w) := \perp$;  \ \ \ \ \  //  the new current $w$-path is $\mathcal{P}_w : w$

$\bullet$     \hspace{68pt}   $gen(w) := w$; \ $low(w) := u$;           \hspace{38pt}  // potential generator  $(u \curvearrowleft w)$

\hspace{8pt} \textbf{1.3} \hspace{5pt} \textbf{else} \ \ \ \ \   // an incoming back-edge

 \hspace{50pt}  $deg(w) := deg(w) - 2$;    \hspace{15pt}  // update $deg(w)$

 \hspace{50pt}        Absorb-subpath$( w, u )$;  \ \ \ \ \ \ \ \ \ \
            // $w$ absorbs a section of the current  $w$-path

\noindent \textbf{end.}\\

\vspace{3pt}
\noindent \textbf{Procedure} \texttt{Gen}-$\mathcal{CS}(w, u, \mathcal{P}_u)$       
\noindent \textbf{begin}

\noindent $\bullet$  \hspace{8pt}    \textbf{if} $( deg(u) = 1 )$ \textbf{then}

\noindent $\bullet$   \hspace{25pt}    $Bridges := Bridges \cup \{ (w, u) \}$;
    \     \hspace{10pt} // add  $(w,u)$ to the 1-cut set

\noindent $\bullet$   \hspace{25pt}       $u := \perp$;    \hspace{110pt} //  eject $u$

  \textbf{else}  \ \ \ \ \ // $deg(u) = 2$

\noindent $\bullet$    \hspace{25pt}
    \textbf{if} ($next(u) = \perp$) \textbf{then}
  \hspace{75pt}  \ \ \   // $\mathcal{P}_u: u$, i.e., the 2-cut involves a \emph{back-edge}.

\noindent $\bullet$  \hspace{40pt}
    \textbf{if} ($bChain(gen(u)) = \perp$) \textbf{then}

\noindent $\bullet$    \hspace{55pt}        $bChain(gen(u)) := (low(gen(u)) \curvearrowleft gen(u)) \oplus (w,u)$;
                                 \ \ \ \ \   // start a \emph{TB}-cut-edge chain $bChain(gen(u))$

                   \hspace{275pt}  \ \ \  with generator $(low(gen(u)) \curvearrowleft gen(u))$

\noindent $\bullet$      \hspace{40pt}
    \textbf{else} $bChain(gen(u)) := bChain(gen(u)) \oplus (w,u)$;    \hspace{40pt} // append $(w,u)$ to $bChain(gen(u))$

\hspace{28pt}
$u : = \perp$;     \hspace{20pt}   //  $\mathcal{P}_u = nil$,  i.e., the $u$-path vanishes after ejecting $u$

\noindent $\bullet$    \hspace{40pt}   \textbf{if} $(lowpt(u) < lowpt(w))$ \textbf{then}
          $gen(w) := gen(u)$;

\noindent $\bullet$     \hspace{25pt}       \textbf{else if} ($tChain(u_1) = \perp$) \textbf{then}


\noindent $\bullet$ \hspace{60pt}
$tChain(u_1) := (u_1, parent(u_1)) \oplus (w,u)$;
        \hspace{15pt}   \ \ \  //start a  \emph{TT}-cut-edge chain $tChain(u_1)$

                   \hspace{240pt}  \ \ \  with  generator $(u_1, parent(u_1))$

\noindent $\bullet$  \hspace{40pt} \textbf{else}   $tChain(u_1) := tChain(u_1) \oplus (w,u)$;
                  \hspace{45pt}    // append $(w,u)$ to  $tChain(u_1)$

\hspace{30pt}
 $u := next(u)$;  \ \ \ \ \     \hspace{20pt}     // eject $u$:   $\mathcal{P}_u := \mathcal{P}_u - u$;

\textbf{output}$(\sigma(u))$;      \hspace{20pt}     // output the 3-edge-connected component $\sigma(u)$

\vspace{3pt}
\noindent \textbf{end.}

\vspace{12pt}
\noindent \textbf{Procedure} \texttt{Absorb-path}$( w )$

\noindent \textbf{begin}

   $x := next(w)$;

  \textbf{while}  $ (x \neq \perp) $  \textbf{do}

\hspace{15pt}     $deg(w) := deg(w) + deg(x) - 2$;

\hspace{15pt}     $\sigma(w) := \sigma(w) \cup \sigma(x)$;

\hspace{15pt}     $x := next(x)$;


\noindent \textbf{end.}

\vspace{12pt}
\noindent \textbf{Procedure} \texttt{Absorb-subpath}$( w, u)$

\noindent \textbf{begin}

  $x := next(w)$;

     \textbf{while}  $( (x \neq \perp) \wedge (dfs(x) \le dfs(u) < dfs(x) + nd(x)) )$  \textbf{do}
    \ \ \ \ \   // $u$ is a descendant of $x$

\hspace{15pt}      $deg(w) := deg(w) + deg(x) - 2$;

\hspace{15pt}    $\sigma(w) := \sigma(w) \cup \sigma(x)$;

\noindent $\bullet$ \hspace{25pt}     $gen(w) := gen(x)$;

\hspace{15pt}       $x := next(x)$;

  $next(w) := x$;


\noindent \textbf{end.}

\noindent\hrulefill


\normalsize
\end{singlespacing}

\begin{lem}\label{CORRECTNESS-cut-edge-chain}
 Let $w \in V$. When the adjacency list of $w$ is completely processed,
  let the $w$-path be $\mathcal{P}_w: w(=w_0) w_1 w_2 \ldots w_k$.

\vspace{-3pt}
\begin{description}
\item[$(i)$]
For each cut-edge chain $\mathcal{E}: e_1e_2 \ldots e_{|\mathcal{E}|}$,
where $e_1 =(y_1 \curvearrowleft x_1)$ or $(x_1 \rightarrow y_1)$, $e_i = (x_i \rightarrow y_i); x_i \prec y_i \preceq x_{i-1}, 2 \le i \le |\mathcal{E}| (|\mathcal{E}| \ge 2)$,  such that
$ x_{\ell + 1} (\text{if exists}) \prec w \preceq x_{\ell} $, for some $\ell \ge 2$,

\vspace{-6pt}
\begin{description}
  \item[$(a)$] If $e_1 = (x_1 \rightarrow y_1)$. then
              $tChain(x_1) = ( (x_1 \rightarrow y_1), (x_2 \rightarrow y_2), \ldots, (x_{\ell} \rightarrow y_{\ell}) )$;

  \item[$(b)$] If $e_1 = (x_1 \curvearrowleft y_1)$. then
              $bChain(y_1) = ( (x_1 \curvearrowright y_1), (x_2 \rightarrow y_2), \ldots, (x_{\ell} \rightarrow y_{\ell}) )$;
\end{description}

\vspace{-6pt}
  \item[$(ii)$]  $gen(w_k) = x$ such that $w \preceq x,$ and
      $(x \curvearrowright low(x)) \in E \backslash E_T$, where  $dfs(low(x)) = lowpt(x) = lowpt(w)$,
   or $(x, low(x)) = (w,w)$.

\end{description}

\end{lem}

\vspace{-9pt}
\noindent \textbf{Proof:}
   By induction on the height of $w$ in $T$.

 At a leaf $w$, since $w \npreceq x_2$ for any cut-edge chain, Condition $(i)$ vacuously holds.

  Since $gen(w) := w$ and $low(w)$ is updated accordingly within Instruction 1.2 whenever the condition $dfs(u) < lowpt(w)$ is detected,
    when the depth-first search back-tracks from $w$,
    $gen(w) = w$ and $low(w) = z$ such that
    $(w \curvearrowright z) \in E \backslash E_T$ and $dfs(z) = lowpt(w)$.
  Condition $(ii)$ thus holds.

   At an internal vertex $w$,
on encountering a vertex $u$ in the adjacency list,
if  $u$ is unvisited, the $\mathit{dfs}$ advances to $u$.
 when the $\mathit{dfs}$ backtracks from $u$,
 if $deg_{\hat{G}_u}(u) = 1$,
 $(w,u)$ is a 1-cut as its removal disconnects $\sigma(u)$ from $G$.
 \textbf{Procedure} \texttt{Gen-CS} is invoked to add $(w,u)$ to the collection of 1-cuts, $Bridges$.

 Let $\mathcal{E}$ be a cut-edge chain whose generator is attached to $T_u$
 (meaning the generator is an edge in $T_u$ or its tail is a vertex of $T_u$)
 such that $x_{\ell + 1} (\text{if exists})  \prec w \preceq x_{\ell} $, for some $\ell \ge 2$.
 Since $(w,u)$ is a 1-cut which is not a cut-edge,
 $w \prec x_{\ell} $
 which implies that $x_{\ell + 1} \prec u \preceq x_{\ell}$.
Since the induction hypothesis applies to $u$,
Condition $(i)$ holds for $\mathcal{E}$.
Moreover, as
 $lowpt(u) = dfs(u) > dfs(w) \ge lowpt(w)$, $gen(w)$ remains unchanged.
  Condition $(ii)$ remains valid.
%

 If $deg_{\hat{G}_u}(u) = 2$,
 \textbf{Procedure} \texttt{Gen-CS} is invoked.

 $(a)$ the $u$-path is $u$, (i.e., $\mathcal{P}_u: u$). Then,
     by Lemma~\ref{2cut-iff-2}$(ii)$, $\{ (w,u), (u \curvearrowright z ) \}$ is the 2-cut,  and
     the corresponding 2-cut in $G$ is $\{ (w,u), (x \curvearrowright z ) \}$, where
     $(x \curvearrowright z )$ is the unique edge such that  $u \preceq x$ and $lowpt(x) = lowpt(u) = dfs(z)$.
     By the induction hypothesis, Condition $(ii)$ holds for $u$ which implies that
     $gen(u) = x'$ such that $u \preceq x'$ and $(x' \curvearrowright low(x')) \in E \backslash E_T$,
      where $dfs(low(x')) = lowpt(x') = lowpt(u)$.
      By the uniqueness of $(x \curvearrowright z )$, $(x' \curvearrowright z ) = (x \curvearrowright z )$ which implies that  $x = x' = gen(u)$ and $z = low(x') = low(gen(u))$.
    Hence, $\{ (low(gen(u)) \curvearrowleft gen(u) ), (u, w) \}$ is the corresponding 2-cut in $G$.
    Clearly, $w \preceq x_2$.
    If $w = x_2$, then $x_2 \prec u$ and $bChain( gen(u) ) = \perp$.
    Therefore,  $\{ (low(gen(u)) \curvearrowleft gen(u) ), (u, w) \}$ is correctly added to $bChain( gen(u) )$.
     If $w \prec x_2$, then $x_{\ell + 1} \prec u \preceq x_{\ell}$ for some $\ell \ge 2$.
     By the induction hypothesis, Condition $(i)(b)$ holds for $\mathcal{E}$ which implies that
     $bChain( gen(u) ) = ( (low(gen(u)) \curvearrowleft gen(u) ), (x_2 \rightarrow y_2), \ldots, (x_{\ell} \rightarrow y_{\ell}) )$.
    Therefore,  $ (w \rightarrow u) = (x_{\ell + 1} \rightarrow y_{\ell + 1})$ is correctly appended to $bChain( gen(u) )$.
Notice that $(low(gen(u)), gen(u) ) \neq (u,u)$ because $lowpt(u) \le dfs(u)$.

If $lowpt(u) < lowpt(w)$, then
   $\mathcal{P}_w : w$ becomes the current $w$-path,
    and $lowpt(w) := lowpt(u)$, $gen(w) := gen(u)$.
   Condition $(ii)$ holds.
If $lowpt(u) \ge lowp(w)$, then as the current $w$-path and $gen(w)$ remain unchanged,
  Condition $(ii)$ continues to hold.

  $(b)$  $\{ (w,u), (u, u_1) \}$ is a 2-cut (i.e., the $u$-path is $\mathcal{P}_u: u u_1 \ldots u_k$, for some $k \ge 1$) in $\hat{G}_u$.
 By Lemma~\ref{2cut-iff-2}$(i)$, the corresponding 2-cut in $G$ is $\{ (u, w), (parent(u_1), u_1) \}$.
    Clearly, $w \preceq x_2$.
    If $w = x_2$, then $x_2 \prec u$ and $tChain( w_1 ) = \perp$.
    Therefore,  $\{(parent(u_1) \rightarrow u_1),(w \rightarrow u)\}$ is correctly added to $tChain( u_1 )$.
     If $w \prec x_2$, then $x_{\ell + 1} \prec u \preceq x_{\ell}$ for some $\ell \ge 2$.
     By the induction hypothesis, Condition $(i)(a)$ holds for $u$ which implies that
     $tChain( u_1 ) = (parent(u_1) \rightarrow u_1  ), (x_2 \rightarrow y_2), \ldots, (x_{\ell} \rightarrow y_{\ell}) )$.
    Therefore,  $ (v \rightarrow w) = (x_{\ell + 1} \rightarrow y_{\ell + 1})$ is correctly appended to $bChain( u_1 )$.

  If $lowpt(w) \le lowpt(u)$,
  then as the current $w$-path remains unchanged, Condition $(ii)$ still holds.
  If $lowpt(u) < lowpt(w)$,
  then
   after ejecting vertex $u$ and including edge $(w,u_1)$, $\mathcal{P}_u$ becomes the current $w$-path and
   $lowpt(w) := lowpt(u)$.
  Since Condition $(ii)$ holds for the $u$-path by the induction hypothesis,
  the condition  holds for the current $w$-path.
  If $lowpt(w) \le lowpt(u)$,
  then as the current $w$-path remains unchanged, Condition $(ii)$ still holds.

If $deg_{\hat{G}_u}(u) > 2$,
the argument that Condition $(i)$ holds is similar to the above  cases, but much simpler.
  If $lowpt(w) \le lowpt(u)$,
  then as the current $w$-path remains unchanged, Condition $(ii)$ still holds.
  If $lowpt(u) < lowpt(w)$,
  then
   after including the edge $(w,u)$, $\mathcal{P}_u$ becomes the current $w$-path and
   $lowpt(w) := lowpt(u)$.
  Since Condition $(ii)$ holds for the $u$-path by the induction hypothesis,
  the condition thus holds for the current $w$-path.

  If $(w,u)$ is an outgoing back-edge of $w$, $(u \curvearrowleft w)$,
  and if $dfs(u) < lowpt(w)$, then
  the current $w$-path becomes $\mathcal{P}_w: w$,  $lowpt(w) := dfs(u)$,
    $gen (w) := w$ and $low(w) := u$, Condition $(ii)$ thus holds.
  If $dfs(u) \ge lowpt(w)$,  as the current $w$-path remains unchanged,
  Condition $(ii)$ continue to hold.
  Since the back-edge does not involve any cut-edge chain satisfying $w \preceq x_{\ell}$, for some $\ell \ge 2$, Condition $(i)$ holds vacuously.

  If $(w,u)$ is an incoming back-edge of $w$, $(w \curvearrowleft u)$,
  \textbf{Procedure} \texttt{Absorb-subpath} is invoked.
  Since the absorption of the section $w_1 w_2 \ldots u$ of $\mathcal{P}_w$ by $w$
  does not affect Condition $(i)$,
   the condition continues to hold.
  Moreover, if $u \neq w_k$, $w$ does not absorb $w_k$.
  Condition $(ii)$ still holds.
  If $u = w_k$, then $w_k$ vanishes, but when the \textbf{while} loop is executed for the last time,
  $gen(w) := gen(w_k)$ ensures that Condition $(ii)$ holds.

When the adjacency list of $w$ is completely processed,
Conditions $(i)$ and $(ii)$ hold for $w$. The lemma thus follows.
    \ \ \ \ \  $\blacksquare$

\vspace{-6pt}
\begin{thm}
\textbf{Algorithm} \texttt{3-edge-connectivity} generates all the 1-cuts, and
all the 2-cuts represented by cut-edge chains for each  of the 2-edge-connected components of
the input graph $G=(V,E)$ in $O(|V|+|E|)$ time.
\end{thm}

\vspace{-9pt}
\noindent \textbf{Proof:}
\textbf{Algorithm} \texttt{3-edge-connectivity}
terminates execution when the adjacency list of the root $r$ is completely processed.
Since $r \preceq x_{|\mathcal{E}|}$ for every cut-edge chain $\mathcal{E}$,
by Lemma~\ref{CORRECTNESS-cut-edge-chain}$(i)$,
all cut-edge chains are correctly computed and
are kept in their corresponding $tChain(x_1)$ or $bChain(y_1)$.

  The time complexity follows from the fact that
   the time complexity of \textbf{Algorithm} \texttt{3-edge-connectivity} of~\cite{T07}  is $O(|V|+|E|)$, and
 the new instructions increase the run time by a constant factor only.
 \ \ \ \ \  $\blacksquare$

\section{Summary}

We presented a linear-time algorithm that  generates the 2-cuts,
the 3-edge-connected components, and the 1-cuts  of a connected undirected multigraph simultaneously  by performing only one depth-first search over the input graph.
The algorithm is conceptually simple and is based on the algorithm of~\cite{T07}.
Our future work is to perform an empirical study of the performances of our algorithm and the algorithm of Georgiadis et al.~\cite{GGIK21} for generating 2-cuts.
Although Georgiadis et al.~\cite{GGIK21} does not address the 3-edge-connected component problem,
their 2-cut algorithm provides a basis for solving the problem:
determine the 2-cuts of the input graph $G$;
 remove the cut-edges from $G$;
for each cut-edge chain with a tree-edge generator, add a new edge~\cite{TWO92};
determine the connected components of the resulting graph which are the 3ecc of $G$.
This multi-pass algorithm should run slower than our one-pass 3ecc algorithm in practice.

\begin{singlespacing}
\begin{small}

\end{small}

\end{singlespacing}
\end{document}